\documentclass[conference]{IEEEtran}
\ifCLASSINFOpdf
 \usepackage[pdftex]{graphicx}
\else
\fi
\begin{document}
%
\title{Rebooting Neuromorphic Design - A Complexity Engineering Approach}

\author{\IEEEauthorblockN{Natesh Ganesh}
\IEEEauthorblockA{App. \& Comp. Mathematics Division, NIST \\
\& Dept. of Physics, CU Boulder \\
Email: natesh.ganesh@colorado.edu}
}


%


\maketitle

\begin{abstract}
As the compute demands for machine learning and artificial intelligence applications continue to grow, neuromorphic hardware has been touted as a potential solution. New emerging devices like memristors, atomic switches, etc have shown tremendous potential to replace CMOS-based circuits but have been hindered by multiple challenges with respect to device variability, stochastic behavior and scalability. In this paper we will introduce a Description$\leftrightarrow$Design framework to analyze past successes in computing, understand current problems and identify solutions moving forward. Engineering systems with these emerging devices might require the modification of both the type of descriptions of learning that we will design for, and the design methodologies we employ in order to realize these new descriptions. We will explore ideas from complexity engineering and analyze the advantages and challenges they offer over traditional approaches to neuromorphic design with novel computing fabrics. A reservoir computing example is used to understand the specific changes that would accompany in moving towards a complexity engineering approach. The time is ideal for a significant reboot of our design methodologies and success will represent a radical shift in how neuromorphic hardware is designed and pave the way for a new paradigm.
\end{abstract}


%
\IEEEpeerreviewmaketitle


\section{Introduction}
The compute resources required to train state of the art (SOTA) machine/deep learning/artificial intelligence (ML/DL/AI) models is increasing at a `super-Moore' rate - doubling every 3.5 months \cite{Openai}, massively increasing the amount of energy required to generate these models \cite{Strubell}. The proposed solutions have been centered around improved co-design around architecture and algorithms as seen in CMOS-based TPUs, FPGAs, ASICs and spike-based hardware. The use of transistors for efficient analog computing has regained some popularity but are not mainstream. There is also growing interest in exploring the use of emerging devices like memristors, photonics and atomic switch networks to build a new generation of AI hardware. While these novel devices show great promise of energy efficiency, high density and non-linearity, they have often been hindered by stochastic device behavior, manufacturing variability and challenges of large scale implementation relative to traditional CMOS. Successful realization of neuromorphic systems with these emerging devices is key to building more efficient hardware to meet the growing demands for compute. 

The goal of this paper is to identify the the fundamental problems in the current framework that hinder the successful integration of these novel devices for AI hardware. If we are able to successfully address these problems, we would then be able to engineer a novel paradigm of complex systems with the potential to realize faster, robust and more efficient information processing \cite{Teuscher}. We will start by analyzing the exponential success we have achieved over the last six decades under a description $\leftrightarrow$ design framework in Sec. II. In Sec. III, we will use the same framework to explain why the time is ideal to completely reboot some of our fundamental ideas in both description and design in order to make progress. Ideas of complexity, complexity engineering and self-organization will be introduced in Sec. IV and V, and will pave the way towards discussing a complexity engineering approach to neuromorphic design in Sec. VI. We will discuss the changes necessary to the descriptive framework with respect to the reservoir computing framework and provide a path forward using non-equilibrium thermodynamics in sec. VII. We will summarize the paper and conclude in Sec. VIII. 

\section{Description $\leftrightarrow$ Design}
The title of this section represents one of the central ideas from this paper. In our field, the \textbf{description} of computation both influences and is influenced by elements of \textbf{design}. The modern computing technology stack is complex with multiple interdependent components. We will break it down to 4 fundamental parts that will be our focus (Fig. 1) - 

\begin{itemize}
    \item[(a)] \textbf{Task} for which the system is being built for.
    \item[(b)] \textbf{Theoretical framework} used to describe how computational states are represented and the algorithm used to achieve the task.
    \item[(c)] \textbf{System architecture} describes how the different parts of the systems are connected.
    \item[(d)] \textbf{Physical computing devices} corresponds to how computation is physically realized i.e. the hardware of the system.
\end{itemize}  

\begin{figure}
\begin{center}
\includegraphics[scale=0.4]{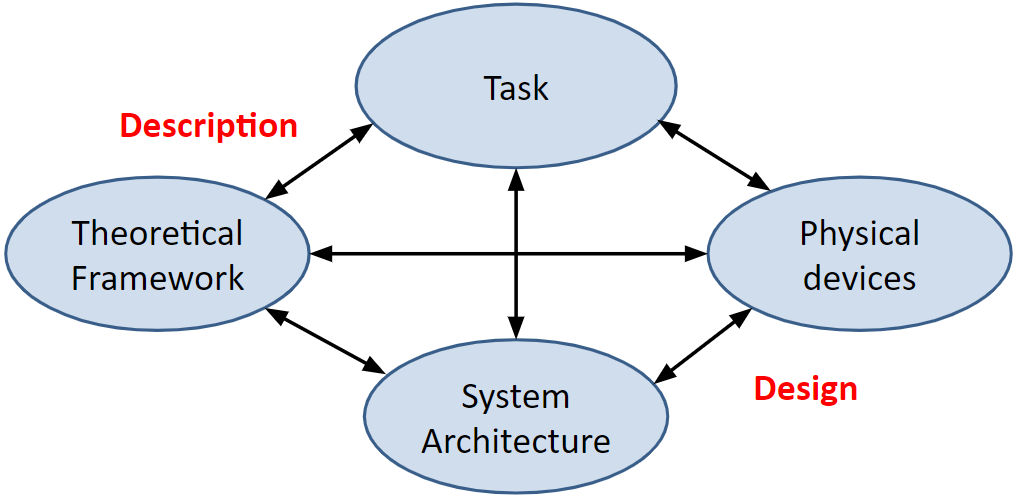}
\end{center}
\caption{Description $\leftrightarrow$ Design - The computing stack divided into four fundamental interconnected components: Description consisting of the Task and the Theoretical Framework, and the Design consisting of System Architecture and Physical Devices.}
\end{figure}

\par\noindent The task and theoretical framework components correspond to \emph{Description} - How is the task described computationally, what is the algorithm, how are inputs and outputs represented? What is considered as achieving the task in a computational manner? The latter two - architecture and devices correspond to \emph{Design} - How are the different blocks necessary to achieve the computation arranged efficiently and what are the physical devices that can realize the specific input and outputs? These 2 categories and the 4 components constantly influence each other, which we will explore further in the next section.

The components during the era of digital computing are -
\begin{itemize}
    \item[(a)] Tasks - Performing large mathematical operations.
    \item[(b)] Theoretical framework - Boolean algebra, finite state automata and Turing machines.
    \item[(c)] System architecture - General purpose computing has been built on a variant of the von Neumann architecture.
    \item[(d)] Physical devices - CMOS devices in binary digital mode.
\end{itemize}

\par\noindent The relative stability of these factors represent a \emph{perfect storm} that drove the digital computing revolution. Let us explore description-design relationship further with respect to these components.

At the heart of modern-day computing is Turing's seminal work in 1936, in which he established a very general model for computation using the idea of Turing machines, showing that \emph{`any function computable by an algorithm, can be computed on a Turing machine'} by manipulating binary symbols like `0' and `1' on the machine tape \cite{Turing1}. Modern computers are not replicas of Turing machines but are based on the idea of manipulating symbols based on efficient algorithms in order to achieve computations. Claude Shannon's work in proving the equivalence between the behavior of networks of electrical switches and Boolean logic functions is another fundamental building block of digital computing \cite{Shannon1}. The first established the theoretical framework and the latter indicated the type of physical systems that can implement the framework - which together pushed for the search for switching devices required to instantiate the binary symbols. 

The primary task of building computers to perform large mathematical calculations was influenced by early digital computers built around the 2nd World War for performing the calculations needed in artillery firing tables, cryptoanalysis, etc and utilized electromechanical switches. The ENIAC machine completed in 1945 utilized vaccuum tubes and is historically important as it introduced the stored-program architecture (also known as the von-Neumann architecture) \cite{Eniac}. It was the first general purpose digital computer, Turing complete and allowed for the system to be reprogrammed by storing the data and program in an external memory. Before the stored-program architecture, we had fixed-program systems in which the program was hardwired in the system for a particular task and could not be reprogrammed - similar to our design of modern day ASICs, albeit a lot less efficient and flexible. Modern day computer architecture is a lot more advanced and complicated, but are built on top of the original von-Neumann architecture.

Transistor technologies (BJT, MOS, CMOS, etc) given their smaller size, faster speeds, lower power consumption, better SNR and ability to be combined with the integrated chip (IC) technology became the preferred device of choice to realize 0's and 1's, and quickly replaced bulkier vacuum tubes in the 1950s. A decade later in 1965, Gordon Moore made his famous observation about the number of transistors on an integrated chip doubling about every two years i.e. Moore's law \cite{Moore}. With the powerful Turing machine theoretical framework, a von-Neumann stored program architecture, Shannon's work on digital switches and the exponential increase in transistor density to realize it, decrease in cost per compute and the growing interest in the scientific study of computers, efficient algorithms, etc, the digital technological revolution was well underway. As the decades passed by, more and more problems across different fields of science, engineering, medicine, economics, etc were made tractable by casting them as a computational problem. And computers became ubiquitous in our everyday lives. Given this exponential progress, it is reasonable to question why the time is right for another revolution of ideas.

\section{Viva la Revolution!!}
The unintended consequence of the incredible success of computing has been a \emph{streetlight effect} i.e. continuing to do what we have already been very successful at. We live in a period where the availability of cheap and powerful compute encourages us to cast all problems in a manner that can be solved by our existing computers, and then look to optimize both the system hardware and software to improve the implementation efficiency. This has also served to discourage a number of ideas to replace conventional systems. 

CMOS devices are considered near irreplaceable in the computing stack with billions of dollars invested in their continued development and in construction of SOTA fabrication facilities. Moore's law has been both the tip of the spear for our progress, and as a shield for CMOS transistor devices (against possible novel replacements) while components (a)-(c) have remained relatively unchanged. Over the many decades, there have been number of research programs focused on identifying devices like spintronics, carbon nanotubes, graphene, quantum-dots, molecular cellular automata, etc (sometimes referred to as unconventional computing \cite{Stepney}). While some of these have been able to match and even surpass CMOS devices in terms of device speed and power dissipation, critics of these novel approaches often point towards their inability to match device robustness, signal-to-noise ratios, scalability and integration with IC design processes. The ability of these devices to construct robust logical gates at scale, which is central to the current computational paradigm is also seen as a major roadblock to their adoption. However with Moore's law slowing down (and Dennard scaling completely stopped) as we approach the physical limits of device scaling, now is the time to invest heavily in the research and development of these new emerging devices at the levels comparable to CMOS technology \cite{Apte}. This should help us both extend our current progress, as well as identify suitable devices for new tasks of interest.
 
The architecture of the system has been the more flexible component when compared to physical devices. FPGAs, ASICs and system on a chip (SoC) for parallel processing, scientific computing, high performance computing, graphic processing units, etc are perfect examples of modifying (c) the system architecture according to the (a) specific task of interest while keeping the fundamental (b) theoretical computing framework (though they use specialized algorithms) and (d) CMOS devices unchanged. Increasingly the focus of the field has shifted away from general purpose computing and towards \emph{AI tasks} - a set of tasks that are associated with intelligence and cognitive abilities. With this shift has come the increasing demand for compute in the field of ML and AI to realize these tasks. The backpropagation algorithm, central to ML was invented in 1986 by Rumelhart and Hinton \cite{Rumelhart}, but the algorithms were not feasible until the availability of GPUs with increased parallelism to perform the large number of computations required \cite{Alexnet}. The lesson here being - the value of an algorithm is dependent on the availability of existing hardware to execute it feasibly. Thus design of computational algorithmic descriptions (of learning and intelligence) are undoubtedly influenced by the type of operations that are feasible on existing hardware i.e. existing \textit{design driving description}.

The hardware solutions to provide the necessary support for ML have mainly focused on architectural improvements, which have been influenced by the machine learning algorithms themselves that needed to be executed. Learning is generally described as weight changes, using gradient descent techniques on a suitable loss function $E$, given by the equation below

\begin{equation}\label{SGD}
    w_{t+1}=w_t-\eta \frac{dE}{dw}
\end{equation}

\par\noindent Learning is achieved during the training phase by performing the above operation in Eq.(\ref{SGD}) on billions of parameters using large amounts of training data. This requires the hardware to perform an extremely large number of matrix multiplication and addition operations. The shift towards more parallel architectures, crossbar structures for more efficient matrix operations, reduced precision arithmetic and improved data movement to combat the memory bottleneck represent significant changes to the system architecture, influenced by descriptions of what learning entails i.e. a case of \textit{description driving design}. These have been adopted by both industry giants (like Intel, NVidia, AMD, ARM, Google) and startups (Cerebras, Mythic, Graphcore, SambaNova) alike to improve the efficiency of the hardware implementing these compute intensive algorithms. Of course more radical descriptions of learning will drive the search for novel hardware (for eg: Shor's algorithm \cite{Shor} for prime factorization was a major driving factor for quantum computing).

While we have focused on the use of transistors as switches for digital computation, they can also function as analog computational elements when used in appropriate device modes (Interestingly the use of transistors in this analog manner exploiting the richer device physics is reminiscent of ideas employed in unconventional computing). An important hardware paradigm that has re-emerged is the field of \emph{neuromorphic computing}. Neuromorphic computing was coined in 1990 by Carver Mead \cite{Mead}, who defined ``neuromorphic systems'' as the use of very large scale integration (VLSI) techniques with analog components that mimicked computation in biological neural systems (and digital blocks for communication). However the use of this term has evolved to become much broader, meaning different things to different researchers. Systems are often defined to be \emph{neuromorphic} at very various levels of the computing stack - algorithm, architecture and device. It includes a wide range of implementations based on both spike-based biologically-inspired algorithms as well deep-learning based artificial neural networks. A detailed survey of neuromorphic systems has been explored in \cite{Schuman} illustrating this very point. Many of these systems have shown tremendous improvements in terms of energy efficiency but much work is needed in improving these algorithms to compete with SOTA deep learning techniques. It might serve the field to clearly define where this neuromorphic boundary lies in order for the term to be meaningful in an useful sense. In any case, hybrid digital-analog systems built based on Mead's original definition can be seen as an natural co-design extension of the fully digital CMOS systems discussed above. In addition to the architectural changes to the system, the transistor devices have been used in an unconventional but natural analog manner to mimic neuronal and synaptic behavior to achieve the tasks in the AI suite. The task, architecture and physical device components have changed to learning tasks, crossbar/parallel architectures and analog computation to efficiently implement the learning algorithms, while the theoretical computing framework i.e. describing learning as the computation of weight changes using Hebbian or gradient descent based techniques, remains consistent across the various systems.

Of particular interest in this paper is the design of neuromorphic hardware using novel emerging devices like memristors \cite{Li}, photonics \cite{Lima}, spintronics \cite{Grollier}, atomic switch networks etc. While we will mainly refer to memristors in this paper (given their increasing popularity in their use in memory, in-memory compute and artificial neural networks), the underlying ideas can be extended to other novel devices as well. Both on-chip and off-chip learning have been achieved in these systems using mainly gradient descent-based algorithms (while some systems have utilized more biologically inspired local Hebbian mechanisms as well). These devices given their small sizes (and thus large density), energy efficiency, speed, non-linearity, etc have shown great promise, but device variability, sneak path currents, self-discharge \cite{Asapu} and latency due necessary control circuitry in dense crossbar structures have hindered their progress with respect to scalability and stackability \cite{Adam}. As incremental advances continue to be made to improve the realization of existing algorithms as well tuning algorithms to account for device variations \cite{Querlioz}, it is necessary to question if the problem isn't one algorithm versus another but rather the underlying computational description and engineering methodologies itself? We must be willing to ask if we need to fully rethink descriptions and design in a manner that maximizes the potential of these novel devices. Changing the description framework alongside the task, architecture and devices will represent a change in all four components concurrently for the first time in over six decades - a big reason why the time might be ripe to make fundamental changes. We will explore this in further detail over the next few sections.

\section{Complex Systems, Complexity Science \& Engineering}
The goal of building neuromorphic hardware is to identify properties of the human brain that are useful for intelligence and emulate it using a different hardware substrate. The human brain is a \emph{complex system}. It is necessary to clearly understand what this term \emph{complex} entails as we look to engineer systems that mimic it. Systems like the human brain, social networks, ant colonies, the internet are a few examples of complex systems. Complexity is roughly defined as being situated between order and disorder. Complex systems are usually identified using some properties that are common to systems that we identify as complex \cite{Mitchell} -
\begin{enumerate}
    \item[(a)] Large number of simple components.
    \item[(b)] Non-linear interaction among parts.
    \item[(c)] No central control.
    \item[(d)] Emergent behaviors like hierarchical organizations, robustness, phase transitions, complex dynamics, information processing capabilities, evolution and learning.
\end{enumerate}

\par\noindent Here emergence corresponds to properties that are seen in the system as a whole but not in the components, that arise due to the interaction between them (colloquially referred to as the `whole being greater than the sum of the parts'). The author in \cite{Mitchell} also distinguishes between disorganized and organized complexity. The first involves billions and billions of parameters and assumes very little interaction between those variables. However our focus will be on the latter, which involves a moderate number of strongly interacting variables and exhibits the properties listed above. The burgeoning science of complexity seeks to identify a unified theory across multiple disciplines. 
  
The main research direction in complexity is to understand it's emergence in natural and engineered systems by identifying and studying the properties of networks. Critical to this task is to define measurable quantities suitable for characterization and analysis \cite{Lloyd1}. An important aspect of this in engineered systems is to address it as a problem that needs to be tackled and to augment the system to cope accordingly \cite{Frei1}. Another option as proposed by the authors in \cite{Buchli} is to engineer systems that looks to take advantage of the complexity rather than suppressing or managing it. This is sometimes referred to as \textit{emergent} \cite{Ulieru} or \textit{complexity engineering} \cite{Wolfram}. It is important to differentiate here between complex and complicated systems. Complicated systems are systems which have a large number of components, predictable, do not show emergent properties and ultimately reducible. Given the difference in properties between complicated and complex systems, the engineering of complex system will look very different to traditional classical engineering ideas and require a significant shift in our design thinking. Let us explore this difference further.

Classical engineering is what is usually taught in universities everywhere and corresponds to applying methods and techniques to solve problems using a reductionist approach, characterized by intuitive analysis, detailed understanding, determinism and predictability \cite{Gershenson}. It requires the systems to be designed to be well-defined and engineers make the reasonable hypothesis that the parts of a system interact in some well-known and well defined ways without further influencing each other. An example of this the \emph{divide and conquer} strategy \cite{Heylighen} - the problem is cut into it's simplest components, and each is analyzed separately, and detailed descriptions are generated. The parts are then connected together as required (ideally the components are modular in nature) and the entire problem is solved. 

Complexity engineering is less formalized currently and is akin to a search of the design space to produce a robust complex system to be situated in a dynamic environment. These systems, by definition are not reducible into their various components. They have \emph{emergent functionality} which means that the required function is not instantiated by a single component or restricted to a part of the system, but instead distributed across the entire system arising at the macroscale due to the interaction of many microscale components (here macro and microscale are relative). It is thus necessary to engineer the right type of interactions between the different components so that the overall system dynamics produces the function of interest. Unlike classical engineering, where the dynamics and functions of every component is fully understood and specified, we will have to relax this constraint in the design of complex systems. We replace it with an approximate understanding of the overall behavior of the system, and the ability to control and predict some aspects of the system output even though it might be difficult (and computationally expensive) to understand how the system produced the output. It is important to understand that it is not a question of which of the two - classical vs complexity engineering is better, but rather a question of the situations for which one might be more suited than the other.

Classical and complexity engineering is also going have different roles for the engineer. Rather than specifying the performance of different components and controlling it, the engineer must now act more as a \emph{facilitator} to guide and enable the system's self-organizing process to produce the results of value, as discussed in \cite{Ulieru}. A loss of complete control and predictability over the systems we design might seem alien for engineers, but it is something we must be willing to explore moving forward. However we are in the early stages in this discipline of complexity engineering. Moving forward we need to expand on the theoretical base from complexity science, a framework to describe and translate concepts such as emergence, evolvability and learning from natural systems to be used in the engineering of technological systems, and a solid methodology to obtain `design' protocols to engineer the complex systems with properties we desire.

\section{Self-Organization - Specification Trade-off}
In addition to being a complex system, the brain (like all biological systems) is a self-organized system. Self-organized systems share some properties that overlap with complex systems like dynamical, robustness, adaptivity and autonomous (with no external or centralized control) \cite{Wolf}. For the purposes of this paper, we will provide a more rigorous definition of self-organization \cite{SelfOrg}. Self-organization is the \textit{`process that produces dissipative non-equilibrium order at macroscopic levels, because of collective, nonlinear interactions between multiple microscopic components. This order is induced by interplay between intrinsic and extrinsic factors, and decays upon removal of the energy source.'} It is not to be confused with self-assembly which is non-dissipative order at equilibrium, persisting without the need for energy. Though we use a more thermodynamics based definition, there are others based on the use of measures that relates self-organization with the increase in statistical complexity of the system \cite{Shalizi}. It is also important in the context of ML/AI to distinguish between self-organization as defined above and self-organizing or Kohonen maps \cite{Kohonen}, which are unsupervised learning algorithms for weights in a neural network. A very useful way to think about classical vs complexity engineering is by using a \textit{self-organization - specification trade-off} introduced in \cite{Buchli}. The author states - \emph{On one hand we need certain functionality in the systems, i.e. we have to be able to specify what the system should do (or a part of it). On the other hand, if we specify ``every'' detail of the system, if we design it by decomposing it, ``linearizing'' the problem, then no self-organization will happen and no emergent phenomena can be harnessed. Thus, there is a trade-off between self-organization on one hand and specification or controllability on the other: If you increase the control over your system you will suppress self-organization capabilities. If you do not suppress the self-organization processes by avoiding constraining and controlling many variables, it is difficult to specify what the system should do''.} 

We can discuss this trade-off in a more concrete manner in terms of the number of variables $N$ used to describe a system at some suitable level of abstraction \cite{Buchli}. Let $N_C$ be the number of constrained variables, which indicate the variables or parameters that the engineer can control in order to extract the required functionality from the system. This makes the rest of the $N$ variables - the unconstrained variables $N_U$ which are not under the control of the engineer and influenced by the system dynamics alone (and evolve as allowed by physical law). By definition we have $N_C+N_U=N$. The two limits include a fully engineered system with $N_C = N$ and no self-organization and no emergent phenomenon on one end, and a fully self-organized system with $N_U = N$. In complexity engineering, the goal is to produce systems with $N_U >> N_C$ - to not exert control over most variables except for a small number of constrained variables to guide the system evolution in the directions that will produce efficient solutions. This does not imply that we can achieve self-organization by taking an existing system and remove the controls in it to make $N_U>>N_C$. Instead we have to take into account the different components and the interactions between them so that the self-organization of the system over time, under the constraints of $N_C$ produces the results we want. By increasing the number of degrees of freedom $N_U$ that self-organize to be larger, we are looking to exploit the intrinsic physics of the system to a greater extent to achieve computing goals. The work presented in this paper can be seen as a natural extension of Mead's originial idea to further exploit CMOS device physics to build analog elements of neuromorphic system, as well as recent work looking to leverage greater amount of physics in algorithms and nanoscale material for neuromorphic computing \cite{Markovic}.

\section{Complexity Engineering Approach to Neuromorphic Design}
The use of complexity engineering approaches to engineer emergence in computing has been limited to unconventional \cite{Polack} and intrinsic \cite{Feldman} computing, and there is no literature of it's use in neuromorphic hardaware to the extent of the author's knowledge. In this section, we will study the advantages of designing neuromorphic hardware using a complexity engineering approach. We start by analyzing neuromorphic design under the traditional engineering approach, which starts with a specific learning task and an algorithm to solve it efficiently. Most learning algorithms - local Hebbian type or global backpropagation based rules are of the form given in Eq.(\ref{SGD}) and operate at the level of weights. This level of the description, that we will refer to as \textit{fine-grained} or \textit{microscale} influences the design of the circuits to implement the algorithm as mentioned earlier. Irrespective of transistor or memristor-based synaptic circuits at the crossbar junctions, we need to build read, write and control circuitry at this microscale level in order to be able to change the weights (constrained variables) - this corresponds to having a system with $N_C>>>N_U$. As we scale up, the additional circuitry required to overcome sneak path currents become increasingly harder and expensive to achieve. Furthermore the issue of variability in the memristor device and behavior is also problematic if we require very specific weight changes. For these reasons, traditional approaches to neuromorphic design using memristors suffer many sizeable challenges. As we continue to make improvements under traditional approaches, we must ask whether it is feasible to engineer such complex systems with the required emergent properties using novel devices in this paradigm \cite{Johnson}. 

We explore the complexity engineering approach by first analyzing a set of conditions a system would need to satisfy in order to be engineered through self-organization \cite{Frei1} - (a) Autonomous and interacting units, (b) Minimal external control, (c) Positive and negative feedback, (d) Fluctuations/variations and (e) Flat dynamic internal architecture. We now map these conditions onto the required neuromorphic hardware systems that is built through self-organization. Such a system will be made of a large number of non-linear neurons interacting through their weights. Currently we build external control into the circuit at a fine-grained (microscale) individual neuron and weight level (i.e. $N_C >> N_U$) to correctly realize the learning algorithm. In order to allow for self-organization, we will have to use a macroscale descriptions of learning (we will explore in further detail in the next section) that would reduce the number of constrained variables and allow the system to evolve freely under it's native dynamics ($N_U>>N_C$), exploiting the rich intinsic behavior of these new devices. Note that by reducing $N_C$, we also make the system evolution more autonomous, satisfying (b) from above. It is also necessary to ensure that as we control the small number of $N_C$, we prevent the system from entering no-go regions in their state-space that are of no value to the users. The use of recurrent architecture and external error signals to influence system evolution can provide the necessary feedback signals to the system. The variations in the device manufacture and behavior are now preferred as we seek for a diversity enabled sweet-spot in self-organized networks \cite{Doyle}. Fluctuations in the microscale (weight) components are tolerable as long as the overall system can realize the necessary functionality and make the system more robust. \emph{Noise} corresponds to unexpected variations to constrained variables $N_C$. On the other hand, variations in the unconstrained variables are acceptable and `noise' is now a resource we can exploit (as done in simulated annealing \cite{annealing}). A flat internal architecture is achieved by using number of neurons of similar behavior and system connectivity that evolves dynamically based on the input stimuli presented to it. For these reasons, neuromorphic hardware systems based on emerging non-linear devices (like memristors) would be a very promising candidate for being engineered through self-organization. These systems can further exploit the existing bottom-up fabrication techniques that are already used to build systems with these devices. It is important to understand the exact type of conditions in which self-organization and complexity engineering approaches will trump traditional ideas. We are not proposing that all neuromorphic systems to be built this way. If we are using digital CMOS devices in which we have engineered away most of the device physics, then traditional techniques will continue to be much better suited.

\section{Macroscale Descriptions of Learning}
A macroscale description of learning corresponds to the change in the theoretical framework that we discussed the need for in section III. For tasks that fall under the AI suite to be realized more effectively, using emerging devices like memristors and brain-inspired architectures requires a change in the description from microscale to macroscale (design driving description), which in turn will also require a change from traditional to complexity engineering methodologies (description driving design). 

In computational neuroscience, Marr's levels of analysis \cite{Marr} was inspired from how information systems have been traditionally built and understood, and there has been a continuous overlap of ideas of between the two fields over the decades. The change in the description of learning can be seen as a technological extension of recent ideas being discussed in computational and systems neuroscience \cite{Blake}. In this paper, the authors describe the classical framework in systems neuroscience as observing and developing a theory of how individual neurons compute and then assembling a circuit-level picture of how the neurons combine their operations. They note how this approach has worked well for simple computations but not for more complicated functions that require a very large number of neurons. They propose to replace it with a deep-learning based framework that shifts the focus from computations realized by individual neurons, to a description comprising of - (a) Objective functions that describe goals of the system, (b) Learning rules that specify the dynamics of neurons and weights and (c) Architectures that constrain how the different units are connected together. They further state that the - \textit{``This optimization framework has an added benefit: as with ANNs, the architectures, learning rules and objective functions of the brain are likely relatively simple and compact, at least in comparison to the list of computations performed by individual neurons.''} This idea of understanding artificial neural networks at a macroscale or coarse-grained level of objective functions, learning rules and architecture of the network overall as opposed to studying the system at the microscale or fine-grained level of individual neurons (where it is often hard to intelligibly describe the system comprising of billions of parameters) has been explored in \cite{Kording}. We are simply suggesting that as we move towards adopting these macroscale descriptions of neural networks, we must look towards complexity engineering methodologies to design and build systems based on these newer descriptions.

\begin{figure}
\begin{center}
\includegraphics[scale=0.7]{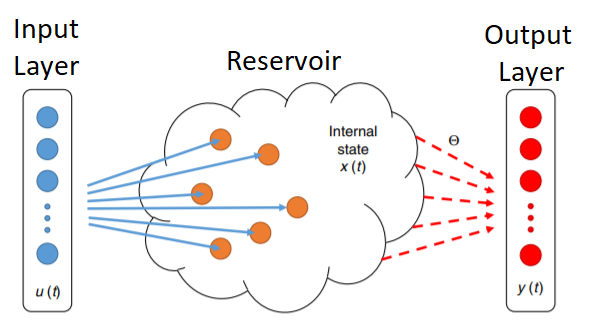}
\end{center}
\caption{Reservoir computing with input layer feeding input signals $u(t)$ to the static reservoir. The reservoir generates a higher order non-linear transformation of the input signals in it's states $x(t)$. The output layer is trained to generate the outputs $y(t)$ using the reservoir states \cite{Du}.}
\end{figure}

We will now introduce a simple example system of reservoir computing (Fig.2) to better clarify the ideas discussed and identify the types of problems that need to be addressed. The reservoir computing paradigm is an umbrella term that includes techniques like liquid state machines and echo state networks \cite{Luko}. They have been physically implemented with non-linear dynamic elements \cite{Tanaka}, \cite{Du}, and provide a much simpler way to train recurrent neural networks (RNN). Reservoir computing systems consists of an input layer, a RNN-based reservoir and a single output layer. The weights in the reservoir remain fixed during training and the network generates a non-linear transformation of the inputs in it's states. The output signal is generated at the output layer as a combination of the reservoir signals. Only the weights in the single output layer are trained using gradient descent with a teacher signal as the target. In order for the system to function properly and approximate the target signal, the weights in the reservoir (generated prior to training using evolutionary algorithms) are chosen such that connection matrix $\mathcal{W}$ of the entire network satisfy the \textit{echo state} or \textit{fading memory property} \cite{Tanaka}. The property is characterized in different ways - a simple and popular one is having the spectral radius of $\mathcal{W}<1$ (a sufficient condition that depends upon input statistics was proposed in \cite{Manjunath}). This is an example of a macroscale condition for learning and inference on the entire reservoir network (that allows for multiple weight level solutions) as opposed to a microscale condition on the individual weights. We can view the static weights as unconstrained variables while the macroscale condition on $\mathcal{W}$'s spectral radius or Schur stability is a constraint. The extension of the static reservoir is an \textit{adaptive reservoir} in which we have a new macroscale reservoir condition to achieve echo state property. This condition accommodates changes to the microscale weights of the network over time depending upon the input statistics in order to adapt and learn new inputs. Unlike traditional RNNs, we do not train the reservoir using an algorithm at the microscale weight-level and instead let the weights evolve (self-organize) as an unconstrained variable. Thus successful identification and implementation of this macroscale condition will result in a RNN with weights that are evolving without external control at the microscale level, while producing the required network functionality. 


We propose a (non-exhaustive) list of properties any macroscale description of learning would need to satisfy in order to be useful in a complexity engineering approach to design. These include the ability to:
\begin{itemize}
    \item[(a)] Address system evolution and self-organization.
    \item[(b)] Be implementation independent like computation. 
    \item[(c)] Quantify information processing and computation. Mapping to existing work in ML is a bonus.
    \item[(d)] Address questions of accuracy and efficiency.
    \item[(e)] Can be studied experimentally and in simulation.
    \item[(f)] Tied to physical law and generate no-go results.
\end{itemize}

\par\noindent The author proposes the use of thermodynamics as a possible macroscale descriptive framework for learning. The field of thermodynamics was invented in the 19th century to address questions of efficiency in engines and has evolved to address the same in information engines \cite{Um}, and computational descriptions, it is universally applicable to all systems. The physical costs of information processing has been widely studied in the thermodynamics of information field \cite{Parrondo}. There is also a rich history of using thermodynamics based ideas in the history of machine learning. Early energy-based models like Hopfield and Boltzmann machines \cite{Rojas}, and free-energy based Helmholtz machines \cite{Dayan} are based on equilibrium thermodynamics - evolving the network towards a state of maximum entropy/minimum free-energy at equilibrium. However self-organized complex systems of interest (like the human brain) are open systems that continuously exchange matter and energy with an external dynamic environment, show a wider range of behavior and are more accurately described by non-equilibrium thermodynamics.

In the last few decades, the field of non-equilibrium thermodynamics has undergone a revolution making a tremendous improvement in theoretical tools available to characterize systems far from equilibrium \cite{Jarzynski}, \cite{Crooks}, \cite{England}. While equilibrium thermodynamics focuses on the distributions of states at static equilibrium (in the infinite time limit), non-equilibrium fluctuation theorems characterize the associated work and heat distributions of dynamical trajectories of states as it is driven by external inputs over finite time. There is a growing body of work in understanding the relationship between non-equilibrium thermodynamics and learning in physical systems \cite{Still}, \cite{Ganesh}, \cite{Ganguli1}, \cite{Ganguli2}. In \cite{Still2}, the author discusses the relationship between minimizing heat dissipation and information-bottleneck based predictive inference in Markov-chain models. There is also very interesting work generalizing the generative model of the Boltzmann machine, in which learning is characterized as distributions of the work done by varying the energy landscape of a non-equilibrium system \cite{Salazar}. The authors in \cite{Salazar} cast the contrastive divergence algorithm for training restricted Boltzmann machines as a physical process that approximately minimizes the difference between entropy variation and average heat, and discuss the relationship between annealed importance sampling \cite{AIS} and the thermodynamic Jarzynski equality \cite{Jarzynski}. In \cite{Boyd}, the authors establish an equivalence between thermodynamics and machine learning by showing that agents that maximize work production also maximize their environmental model’s log-likelihood. The relationships between computational descriptions of learning processes and non-equilibrium thermodynamic descriptions of the same as a physical process are what we are looking to achieve. These descriptions can serve as the basis for developing non-equilibrium control protocols using ideas from thermodynamic control \cite{Deffner}, \cite{Rotskoff} to design physical systems to realize these protocols. These systems will realize the corresponding thermodynamic condition and by the aforementioned equivalency the computational learning as well. Engineering hardware based on these thermodynamic descriptions might seem alien when compared to our existing top-down standard protocols for design using computational descriptions. However designing physical systems based on thermodynamic considerations is very common in the field of molecular and bio-engineering, and unconventional computing. A good example of this are chemical Boltzmann machines \cite{Poole}, in which the authors construct a mapping from Boltzmann machines to chemical reactions. And these reactions are usually designed based on kinetic and thermodynamic factors \cite{Prausnitz}.

\begin{figure}
\begin{center}
\label{ASNs}
\includegraphics[scale=0.42]{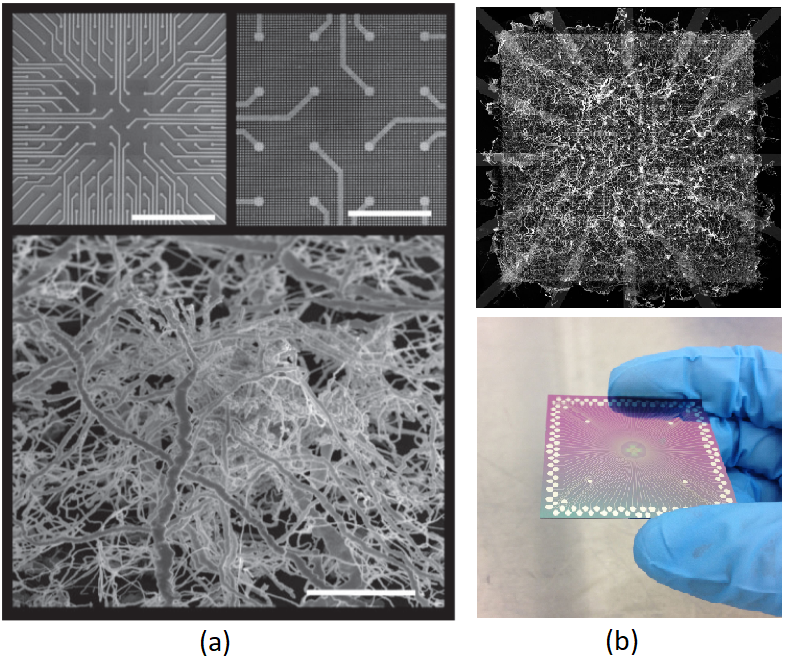}
\end{center}
\caption{(a) An optical image of an SiO2 coated wafer with 120 Pt electrodes (top) \& SEM of a self-assembled Ag+ network after submersion in AgNO3 (bottom). (b) The silver nanowire network (top) takes the form of a tiny square of mesh at the center of the device (bottom) \cite{Steig}.}
\end{figure}

A list of some of the considerations when trying to engineer a self-organized system is provided in \cite{Fiakowlski} - (a) identify suitable interactions, (b) choosing competing interactions and potentials, (c) choosing proper scale and (d) synthesis - moving from small systems with minimal components to larger systems. A detailed review of the principles behind directed self-organization by manipulating the energy and entropy landscape of the system is available in \cite{Furst}. This change in our design approach to computing hardware is going to need a restructuring of our philosophies and significant interdisciplinary work. We do not have start from scratch though, as we can look to build upon work in nanoarchitectonics \cite{Ariga}, computational matter and in-materio evolution \cite{Konkoli}. We also have a good idea of the properties that we want in the final engineered self-organized systems - a network with large number of heterogeneous components, sparse connections with synaptic behavior, scale-free/small-world topology, criticality, etc. Examples of self-organized networks that satisfy such properties and have been shown to be capable of achieving learning include \cite{Steig} (Fig.3), \cite{Alvarez}, \cite{Manning}, \cite{Bose} and \cite{Milano}. Fabrication of these networks are not based on computational descriptions and provide the ideal base to experiment on and build a framework to understand novel macroscale descriptions of learning and the relationship between choices in the design process to functional capabilities of the self-organized system.

\section{Discussion \& Conclusion}
In this paper, we studied the connections between the Description and Design components of the computing stack. This framework allowed us to understand the exponential success that we have achieved over the years, and also identify the issues facing us moving forward in the design of neuromorphic hardware. If our goal is to engineer energy efficient neuromorphic hardware that can mimic the abilities of a complex self-organized brain using emerging stochastic devices, we must be willing to replace the traditional reductionist engineering approach with complexity engineering methodologies. This will require a significant shift in how we understand computation and describe learning in physical systems, our design philosophy on what it means for a complex system to be `designed,' and the role of the engineer in these systems. In this paper, the author identifies two challenges that need to be addressed to achieve progress - (a) identifying new macroscale descriptions of learning that would leverage the intrinsic physics of the system to a greater degree and suggested non-equilibrium thermodynamics as a possible path forward and (b) take inspiration from other fields to develop new design protocols using the above descriptions that would have better synergy with these emerging fabrics.

The author recognizes the tremendous challenges and work that lies ahead of us, but views these as an unique set of opportunities that not are often available to the research community. Complexity engineering is a relatively new field that requires a lot of research to be formalized, expanded and brought into the mainstream of hardware design. One can look back at history and point to traditional hardware design to be facing the exact same challenges at the start of the digital computing paradigm in the 1940s. However eight decades later, with the appropriate investments in research we have made great strides in understanding traditional design and building a large number of tools and infrastructure to make it feasible and profitable. The author hopes that given the massive benefits that we could reap from efficient self-organizing hardware for AI applications, the community will increase focus on these new ideas. Success would allow us to meet the growing compute demand in the short term and usher in a technological revolution in the long term.

\bibliographystyle{ACM-Reference-Format}

\end{document}